\documentclass[twocolumn, amsmath,superscriptaddress, amsfonts,prb]{revtex4}
\usepackage{graphicx}
\usepackage{url}
\begin{document}

\title{Understanding the movements of metal whiskers }
\author{V. G. Karpov}\email{victor.karpov@utoledo.edu}\affiliation{Department of Physics and Astronomy, University of Toledo, Toledo, OH 43606, USA}
\begin{abstract}
Metal whiskers often grow across leads of electric equipment causing short circuits and raising significant reliability issues. Their nature remains a mystery after several decades of research. It was observed that metal whiskers exhibit large amplitude movements under gentle air flow or, according to some testimonies, without obvious stimuli. Understanding the physics behind that movements would give additional insights into the nature of metal whiskers. Here, we quantitatively analyze possible mechanisms of the observed movements: (1) minute air currents, (2) Brownian motion due to random bombardments with the air molecules, (3) mechanically caused movements, such as (a) transmitted external  vibrations, and (b) torque exerted due to material propagation along curved whiskers (the garden hose instability), (4) time dependent electric fields due to diffusion of ions, and (5) non-equilibrium electric fields making it possible for {\it some} whiskers to move. For all these mechanisms we provide numerical estimates. Our conclusion is that the observed movements are likely due to the air currents or electric recharging caused by external light or similar factors.

\end{abstract}

\date{\today}

\maketitle
\section{Introduction}\label{sec:intro}
Metal whiskers present a significant factor behind the reliability of many electronic devices. In particular, tin whiskers threaten many technologies  using Sn based solder alloys. The mechanisms of metal whisker development and their properties remain insufficiently understood after almost 70 years of research. \cite{bibl,brusse2002,davy2014,zhang2004,tu2005,bunian2013,karpov2014,karpov2015} Several proposed models of whisker nucleation and growth remain insufficiently quantitative and/or debatable. \cite{karpov2014,karpov2015, sobiech2008,sobiech2009,osenbach2009,yang2008,cheng2011,su2011,buchovecky2009,pei2012}
\begin{figure}[ht]
\includegraphics[width=0.47\textwidth]{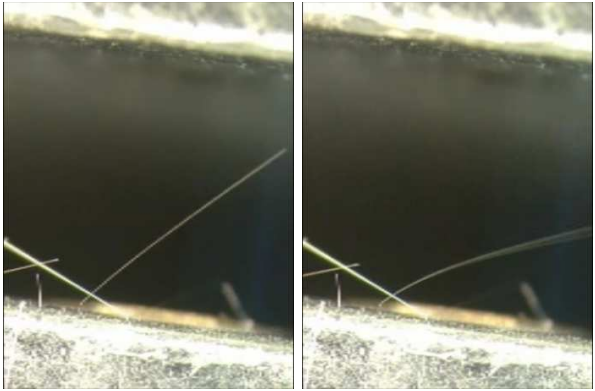}
\caption{Snapshots of $\sim 2.5$ mm long tin whisker separated by time interval of $\sim 0.05$ s. The whisker is moved by a minute air current created by a gentle expiration towards the whisker. Courtesy of the NASA Electronic Parts and Packaging (NEPP) Program, \cite{video}
{\it http://nepp.nasa.gov/whisker}.\label{Fig:cartoon}}
\end{figure}

Here we discuss one of the long known yet poorly understood whisker properties: their ability to exercise movements of significant amplitudes comparable to whisker lengths. These movements have been observed multiple times in response to rather gentle (say, expiratory) air flows, \cite{video,video1} as illustrated in Fig. \ref{Fig:cartoon}.

More intriguing is that many researchers have reported spontaneous (unintended) whisker movements taking place without any obvious stimuli. The author too has observed unintended movements of {\it some} Zn whiskers using a magnifying glass and a LED flash light (whisker infested Zn samples, courtesy of J. Brusse, the NASA NEPP Program); however the question of reliability of those observations or possible uncontrolled air flow effects remains open.

The testimonies of `spontaneous' metal whisker movements have been a topic for extensive discussions in the metal whisker community, including through the `tin whisker group teleconference',\cite{teleconf} which resulted in various hypotheses, such as, e. g. `Brownian movements', `mechanical vibrations', or `garden hose instability', and some others. These hypotheses about metal whisker movements, either air current stimulated or spontaneous, have not been quantitatively tested.

This paper provides a so far missing quantitative analysis of whisker movement mechanisms. Its results below are accurate to the order of magnitude. This approximate approach appears adequate in view of various uncertainties related to whisker parameters, as well as the state of the art approximate nature of theoretical descriptions of stochastic and hydrodynamic phenomena involved. The latter issues make exact model solutions excessive and beyond any reasonable accuracy.

Assuming a diverse readership including many non-physicists, a comment may be in order regarding the nature of the order-of-magnitude approximate analysis.  It is well known with many subjects in the hydrodynamic theory, statistical physics, the theory of disordered systems, and many other subjects where exact solutions are impossible or useless. An important lesson learned  from that multitude of applications, is that the order-of-magnitude approach provided results that are generally good to the accuracy of insignificant numerical factors, such as, say, 0.7, or 2.1, etc., as compared to the exact results where available. In the meantime, it is more intuitive, fast, and economical avoiding unnecessary mathematical complications.

Following the standard notations, the equality sign $=$ will be everywhere replaced with the `equal in-the-order-of magnitude' sign $\sim$, meaning `to the accuracy of a numerical factor'. For example, the area of a circle of diameter $d$ is estimated as $A\sim d^2$, instead of the exact $A=\pi d^2/4$, different by a numerical factor of $\pi /4\approx 0.75$. Below, we will compare, at some points, the `order-of-magnitude' and exact results.

In addition, the `order-of-magnitude' analyzes often use intuitive estimates for derivatives and integrals, such as $\partial y/\partial x\sim \Delta y/\Delta x$ and $\int _a^{a+\Delta X}y(x)dx\sim y(a)\Delta x$ where $\Delta y$ and $\Delta x$ are the intervals for variables $x$ and $y$. These approximations generate additional numerical uncertainties. For example, assuming $y=x^2$, the latter estimates yield, $\partial y/\partial x\sim x^2/x=x$ and $\int _0^xydx\sim yx=x^3$, missing respectively the numerical factors of 2 and 1/3, insignificant as not changing the order of magnitude of the estimated quantities. Many useful examples of such approximations are described in the classical text by Migdal. \cite{migdal}

The observed insignificance of numerical factors, typically not too different from unity, has found a commonly shared conceptual explanation (folklore attributed to Einstein) that, because numerical uncertainties at different steps in complex processes are mutually independent, they tend to cancel each other similar to random displacements in diffusion processes (i. e. `errors propagate by diffusion'). Say, the coefficients of 2 and 1/3 in the above examples, can combine into 2/3, which is even closer to unity. There are very few known exceptions where such cancelation does not take place. This amazing observation is fully consistent with the fact that the laws of physics typically do not include any significant numerical multipliers when written a `natural', say, Gaussian units (as opposed to SI where such multipliers are enforced).

Using the `order of magnitude'analysis throughout multiple subjects in this paper will offer a practicum that can be helpful to graduate students and other audience interested in practical approximate techniques of theoretical physics. In addition, this paper provides a brief tutorial on general bending dynamic properties of elastic rods (modeling metal whiskers); they are not often taught in the standard physics curriculum and can be useful for researchers interested in metal whisker effects.

The paper deals with multiple independent subjects; hence, quite a few quantities denoted with various letters. The intent was to keep the notations close to standard, such as, e. g. $A$ for area, $D$ for diffusion coefficient, etc. One obviously inconvenient case is $d$ for the diameter, which meaning could be mixed with that of differential. To avoid any confusion, $d$ is never used as a differential in what follows; in several cases where the notion of differential becomes unavoidable, we use the partial derivative  denoted with $\partial$.

The paper is organized as follows. Section \ref{sec:model} describes the elastic beam model of a metal whisker, and especially its bending properties for uniform and nonuniform whiskers. Sec. \ref{sec:air} provides estimates of possible whisker movements due to viscous air flow. In Sec. \ref{sec:brown}, the analysis of Brownian movements is done for metal whiskers in comparison with the standard Brownian movement of small particles, as well as the viscous drag effects. Possible mechanical causes of whisker movements are analyzed in Sec. \ref{sec:mech}. Finally, Sec. \ref{sec:electric} discusses the possibility of whisker movements due to electrostatic effects. The numerical estimates corresponding to all the analyzed mechanisms are summarized in Sec. \ref{sec:num}. The paper conclusions are presented in Sec. \ref{sec:concl}.

\section{Elastic beam whisker model}\label{sec:model}
\begin{figure}[ht]
\includegraphics[width=0.45\textwidth]{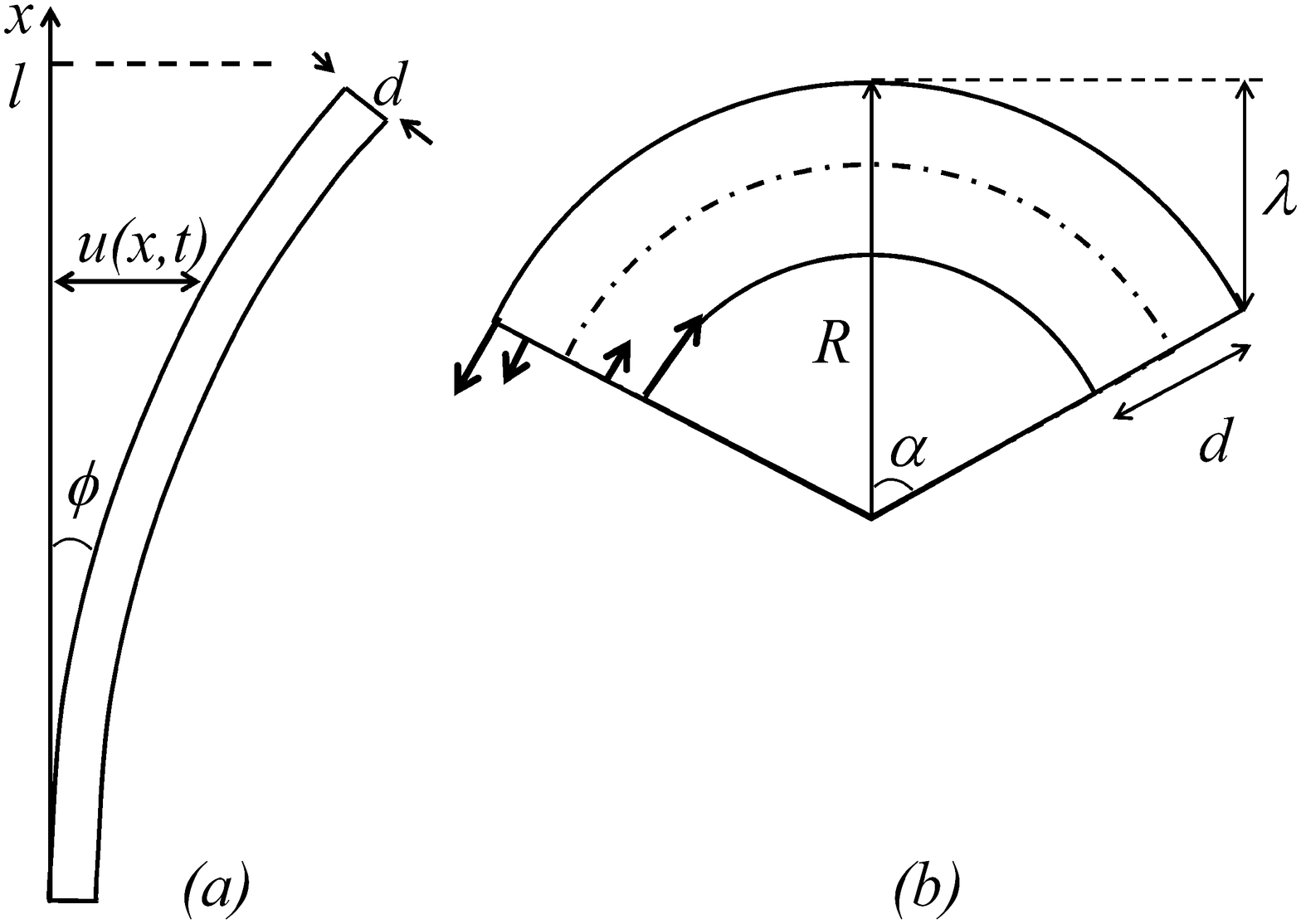}
\caption{(a) Elastic beam of length $l$ and diameter $d$ characterized by the deflection $u(x,t)$ from axis $x$. (b) A beam bent with the curvature radius $R$ and maximum deflection $[u(x,t)]_{\rm max}=\lambda$ (not to scale $d\ll l$). Fat tangential arrows show stresses that have opposite directions above and below the neutral line presented with a dash dot line.\label{Fig:whiskgeom}}
\end{figure}

Because the observed whisker movements are reversible, the elastic approximation appears to be adequate. We describe each whisker as a cylinder shaped elastic beam of a relatively small diameter, $d\ll l$, where $l$ is the whisker length, parallel to the $x$-axis of a Cartesian coordinate system (Fig. \ref{Fig:whiskgeom}) (a)), and limit ourselves to bending deformations.

\subsection{Euler-Bernoulli equation}\label{sec:EB}
The elastic beam deformations are described in terms of deflection $u(x,t)$ by the Euler–Bernoulli equation (see e. g. Refs. \onlinecite{leissa2011,boresi1993} or dozens of other excellent sources)
\begin{equation}\label{eq:EB}
YI\frac{\partial ^4u}{\partial x^4}+\rho A\frac{\partial ^2 u}{\partial t^2}+B\frac{\partial u}{\partial t}=f(x,t)
\end{equation}
where $Y$ is the Young's modulus, $I$ is the polar area moment of inertia of the cross-section (defined as the integral of the square of a radius vector of a typical point in the cross-section over the cross-section area),
$\rho $ is mass per unit volume constant throughout the cross-section, $A$ is the cross-sectional area, and $B$ is the friction force coefficient, and $f(x,t)$ is the distributed force per length that depends on the coordinate and time.

\subsection{Qualitative description}\label{sec:qual}

While Eq. (\ref{eq:EB}) should describe the entire beam dynamics, in what follows, we rely on an approximate `order-of-magnitude' approach facilitating its understanding, and neglecting possible quantitative discrepancies in numerical coefficients.

\subsubsection{Uniform whiskers}\label{sec:unif}
Here, we consider whiskers with uniform cross-section along the longitudinal coordinate. We start with estimating the beam deflection $\lambda$ in response to force $F$, either distributed or exerted close to the beam edge. Consider an ark of length $l$ illustrated in Fig. \ref{Fig:whiskgeom} (b). Its upper and lower parts are respectively stretched and squeezed due to bending. A line between them, called the neutral axis, is neither stretched no squeezed maintaining the same length as before deformation. Denoting the arc curvature radius by $R$ and using the geometry in Fig. \ref{Fig:whiskgeom}, it is straightforward to estimate the maximum deflection as
\begin{equation}\label{eq:lambda}\lambda\sim l^2/R.\end{equation}

The length of arc at distance $d$ above (below) the neutral line is $(R\pm d )\alpha$ where $\alpha$ is the arc angle. Therefore, the relative deformations of the upper and lower parts are estimated as $\pm \Delta l/l\sim \pm d/R$. The corresponding stresses are $s=Y\Delta l/l\sim \pm Yd/R$ where $Y$ is the Young's modulus. They result in the torque density $\pm sd$ (i. e. force density $s$ times lever $d$), and the integral torque $M\sim Asd\sim Yd^4/R$ where $A\sim d^2$ is the cross sectional area.  Expressing $R$ from Eq. (\ref{eq:lambda}), one gets the torque related to the bending deformation,
\begin{equation}\label{eq:torque}
M\sim Y\frac{\lambda d^4}{l^2}.
\end{equation}

On the other hand, the torque can be expressed through a more or less perpendicular exerted force $F$ as $M\sim Fl$. Comparing the latter to Eq. (\ref{eq:torque}) yields,
\begin{equation}\label{eq:deflection}
\lambda\sim \frac{Fl^3}{Yd^4}\sim \frac{F}{K}\left(\frac{l}{d}\right)^2\equiv \frac{F}{K_b}.\end{equation}
Here, we have introduced the compressive spring constant $K=Yd^2/l$ as the proportionality coefficient between the force $sA$ and compressive elastic deformation $\Delta l$, and the effective bending spring constant $K_b=K(d/l)^2$.

As an illustration of the `order-of-magnitude' approach, the exact result \cite{leissa2011} is that force $F$ acting perpendicularly on the end of a cantilever beam creates a displacement given by Eq. (\ref{eq:deflection}) with a numerical coefficient of 1/3, while the case of uniformly distributed force yields the coefficient 1/6.

Eq. (\ref{eq:deflection}) shows that bending deformation $\lambda$ is by the factor $(l/d)^2\gg 1$ greater than the compressive deformation $F/K$. In other words, the bending elastic spring $K_b$ is by the factor $(d/l)^2\ll 1$ smaller than the compressive spring constant $K$.

The above simple estimates lead to the characteristic frequency of transverse vibrations,
\begin{equation}\label{eq:freq}
\omega _b\sim\sqrt{\frac{K_b}{m}}\sim\frac{d}{l^2}\sqrt{\frac{Y}{\rho}}\sim\frac{d}{l}\omega,
\quad\omega\equiv\sqrt{\frac{K}{m}}\sim\frac{1}{l}\sqrt{\frac{Y}{\rho}}.
\end{equation}
where $m\sim \rho ld^2$ is the beam mass. We observe that bending vibrations have frequencies by the aspect ratio $d/l\ll 1$ lower than the standard elastic vibration frequencies $\omega$.

Estimates in Eq. (\ref{eq:freq}) can be verified in several ways. For example, by setting equal the terms on the left-hand-side of Eq. (\ref{eq:EB}) and estimating $\partial ^4u/\partial x^4\sim u/x^4$ and $\partial ^2u/\partial t^2\sim u\omega _b ^2$. Or by considering the standard angular dynamics via $I_0\partial ^2\phi /\partial t^2=M$ where
\begin{equation}\label{eq:momin}I_0\sim ml^2\sim \rho l^3d^2\end{equation} is the moment of inertia, $\phi $ is the deflection angle,  and $M$ is given by Eq. (\ref{eq:torque}) with $\lambda \sim l\phi$.

To further illustrate the accuracy of the `order-of-magnitude' treatment, note that the vibrational spectra of cantilever beams are described as, \cite{leissa2011}
\begin{equation}\label{eq:exactfr}
\omega _b=\beta ^2\frac{1}{l^2}\sqrt{\frac{YI}{\rho A}}
\end{equation}
where $\beta ^2$ is a numerical factor labeling different vibrational modes (in terms of the number of nodes). Taking into account that $I\approx A^2$, Eq. (\ref{eq:exactfr}) coincides with Eq. (\ref{eq:freq}) to a numerical factor $\beta ^2$, which for the first mode is close to one. The coefficient $\beta$ increases with the mode number not accounted for by the approximate Eq. (\ref{eq:freq}). However, the existence of higher excited vibrational modes is irrelevant to this paper subject not dealing with resonance phenomena.

Since \begin{equation}\label{eq:vs}\sqrt{\frac{Y}{\rho}}\sim v_s\end{equation}
represents the sound velocity $v_s$, one can conclude that $\omega$ is the characteristic reciprocal time, over which the system retains a coherent translational dynamics (all parts of the body move coherently). $\omega _b$ represents a similar value for the bending dynamics.

\begin{figure}[b!]
\includegraphics[width=0.57\textwidth]{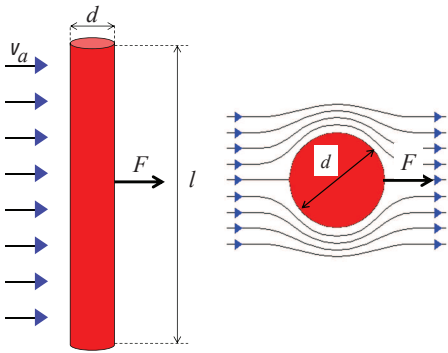}
\caption{Side and top views of an airflow around a long cylinder shaped whisker with arrows showing the current lines. $F$ is the integral force on the cylinder. \label{Fig:wind}}
\end{figure}

The latter can be shown more explicitly by approximating $F$ in Eq. (\ref{eq:deflection}) through the Newton's second law as
\begin{equation}\label{eq:Newton}F\sim m\lambda /\tau _b ^2\quad {\rm with}\quad m\sim \rho d^2l\end{equation}
where $\tau _b$ is the deflection time and the acceleration is estimated as $\lambda /\tau _b ^2$.  Substituting Eq. (\ref{eq:Newton}) into Eq. (\ref{eq:deflection}) yields,
\begin{equation}\label{eq:taulambda}
\tau _b\sim \frac{d
}{v_s}\left(\frac{l}{d}\right)^2\sim\frac{1}{\omega _b}.\end{equation}

The estimates in Eqs. (\ref{eq:vs}) and (\ref{eq:Newton}) are limited to dynamics governed by the whisker bare mass $m$. They do not apply to the case of viscous drag driven whiskers, particularly in Brownian motion, where the dynamic properties are determined by a composite system of a whisker and a fluid (see Sec. \ref{sec:brown}.)

Consider next the viscous drag on a cylinder whisker with high aspect ratio $l/d\gg 1$ that corresponds to the third term on the left hand side in Eq. (\ref{eq:EB}). Because $d$ is the only characteristic length at hands,  one concludes that the air velocity field $v_a({\bf r})$ is significantly disturbed at distances $r\lesssim d$ from the whisker; hence, $|\nabla v_a|\sim v/d$ where $v$ is the whisker velocity. As a result, to  the logarithmic accuracy, \cite{lee1986,vandyke1975} the viscous drag is estimated as
\begin{equation}\label{eq:viscdrag}F\sim \eta A_s|\nabla v_a|\sim \eta vl\end{equation}
where $A_s\sim ld$ is the whisker side area.

\begin{figure}[t!]
\includegraphics[width=0.18\textwidth]{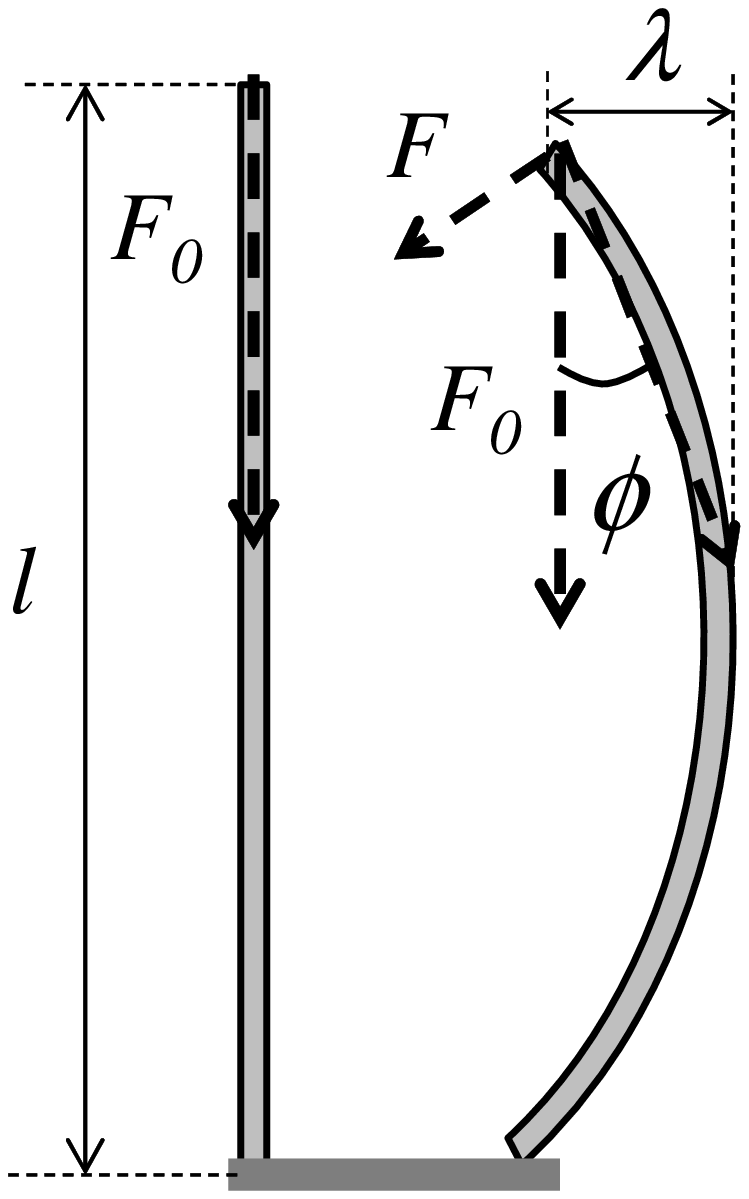}
\caption{Buckling instability results in switching from vertical beam purely compressed under force $F_0$ (left) to a bent configuration characterized by a finite deflection $\lambda$ (right). \label{Fig:buckling}}
\end{figure}

For completeness, we briefly describe also the condition of buckling instability. It represents switching from a purely compressed (no beam deflection) to a laterally-deformed state characterized by a finite deflection $\lambda$ illustrated in Fig. \ref{Fig:buckling}. The switching between these states takes place when the compressive force $F_0$ exceeds a certain critical value $F_{0c}$ that depends on the beam length $l$, or, in other words, when, the beam length $l$ exceeds a certain critical values $l_c$ that depends on $F_0$.

To derive the switching condition, we note that the bending force $F$ that appears in Eq. (\ref{eq:deflection}) can be expressed through the originally applied compressive force $F_0$ as $F=F_0\sin\phi$ where $\phi$ is the bending angle shown in Fig. \ref{Fig:buckling}. Estimating $\phi\sim \lambda/l$ and substituting $F=F_{0}\lambda /l$ in Eq. (\ref{eq:deflection}) yields,
\begin{equation}\label{eq:critF}
F_0=F_{0c}\sim Yd^4/l^2.\end{equation}
When $F_0<F_{0c}$, Eq. (\ref{eq:deflection}) cannot be satisfied. On the other hand, when $F_0>F_{0c}$, the above estimate for the angle becomes inaccurate and we are left with Eq. (\ref{eq:deflection}).

Expressing the beam length through Eq. (\ref{eq:critF}), gives the critical beam length,
\begin{equation}\label{eq:critl}
l_c=\sqrt{Yd^4/F_0}.\end{equation}
The above described buckling yields a significant deflection $\lambda\lesssim l$. The results in Eqs. (\ref{eq:critF}) and (\ref{eq:critl}) coincide with that of exact analysis \cite{leissa2011,boresi1993} to the accuracy of numerical coefficients of the order of unity.

\subsubsection{Nonuniform whiskers}\label{sec:nonunif}
\begin{figure}[b!]
\includegraphics[width=0.28\textwidth]{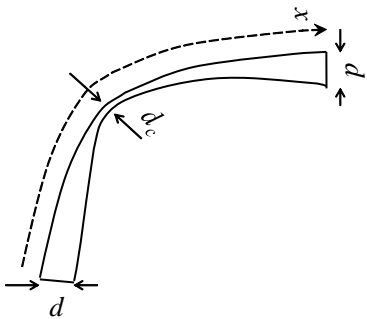}
\caption{Sketch of a whisker with constriction where the diameter $d_c$ is much lower than its average diameter $d$. \label{Fig:constr}}
\end{figure}

Consider next whiskers with significantly nonuniform cross-sections i. e. with diameters $d(x)$ significantly varying along the whisker longitudinal coordinate $x$, as illustrated in Fig. \ref{Fig:constr}. Such a whisker can be thought of as a set of small domains, such that $d(x)$ in each of them is more or less uniform. In static equilibrium, the forces on subsequent domains must be equal. Therefore, Eq. (\ref{eq:deflection}) predicts then that the bending angles in individual domains are inversely proportional to $d^4$, i. e. a whisker integral bending is determined by its constriction region (if any) where the diameter $d_c\ll d$.

More specifically, taking the individual domain lengths $\delta l_i=d_i\equiv d(x_i)$, setting subsequent domain forces equal and using Eq. (\ref{eq:deflection}) yields
\begin{equation}\label{eq:constr}\delta \theta _i \sim\frac{\lambda _i}{\delta l_i}\propto \frac{1}{d_i^2}.\end{equation}
Eq. (\ref{eq:constr}) can be further generalized to give the integral bending angle,
\begin{equation}\label{eq:thetaint}
\theta _{\rm int}\sim \frac{F}{Y}\sum _{i}\frac{p_i}{d_i^2}\end{equation}
where $p(d_i)$ is the probability to find a domain of diameter $d_i$, so that $\sum _ip(d_i)=1$.

As an example, consider a whisker of length $l=1$ mm and average diameter $d_{\rm av}=10$ $\mu$m having a constriction with diameter $d_c=1 $ $\mu$m extending over distance $\delta l\sim d_{\rm av}$ corresponding to the sketch in Fig. \ref{Fig:constr}.  In the constriction region, we set $\delta l_i= 1$ $\mu$m which makes $n_c\sim 10$ domains forming the constriction and yielding $p(d_c)\sim d_{\rm av}/l\sim 0.01$. On the other hand, the average diameter region will contain $n_{\rm av}\sim l/d_{\rm av}=10^2$ domains of length $\delta l_i=d=10$ $\mu$m each. According to Eq. (\ref{eq:thetaint}), each domain in the constricted region will contribute the deflection angle $\theta _i$ that is by the factor $(d_{\rm av}/d_c)^2=100$ greater than that by the average diameter domain. Therefore, the relatively short constriction region (of $\sim 1$ \% of the total length) will contribute to the integral bending, approximately the same angle as the rest 99 \% of the whisker, as dictated by the relation $n_c/d_c^2\approx n_{\rm av}/d_{\rm av}^2$.

It is obvious that constrictions will affect whisker vibrations. This effect can be readily described for the case of a distinct local constriction of the type presented in Fig. \ref{Fig:constr}.  The constriction related bending elasticity, $K_{bc}$ is softer than that of the average one,
\begin{equation}\label{eq:constK}
K_{bc}\sim \frac{l}{l_c}\frac{d_c^2}{d_{\rm av}^2}K_b\sim \frac{ld_c^2}{d_{\rm av}^3}K_b\ll K_b,\end{equation}
while the vibrating mass $m$ is more or less the same. As a result, the characteristic vibrational frequency corresponding to a vibrating whisker with constriction is estimated as
\begin{equation}
\omega _{bc}\sim \omega _b\sqrt{\frac{ld_c^2}{d_{\rm av}^3}}K_b\ll \omega _b,\end{equation}
and the bending time $\tau _{bc}$ is by the same factor longer than that ($\tau _b$) without the constriction.

%We shall end this section with some numerical estimates using $d\sim 10$ $\mu$m, $l/d\sim 10^2-10^3$, $v_s\sim 3\cdot 10^5$ cm/s, and $\rho\sim 10$ g/cm$^3$. This yields $\omega \sim 3\cdot (10^5-10^6)$ 1/s and $\omega _b\sim 3\cdot (10^2-10^4)$ 1/s. Note that the observed whisker deflection times $\tau _b\lesssim 0.1$ s through Eq. (\ref{eq:taulambda}) are consistent with the latter estimate of $\omega _b$.

\section{Air flow effects}\label{sec:air}
Here we test the hypothesis of minute current flows causing  the observed whisker movements. The fact that whisker deflections $\lambda$  can be significant, is readily quantified by using Eqs. (\ref{eq:deflection}) and (\ref{eq:viscdrag}). Combining them and setting $\lambda\sim l$ results in the following estimate for the local air velocity that produces significant (comparable to the whisker length) deflections,
\begin{equation}\label{eq:wind}
v_a\sim \frac{Yd^4}{\eta l^3}.\end{equation}

The numerical estimates in Sec. \ref{sec:num} show that thin (micron in diameter, or constricted to such neck diameter) and simultaneously tall ($l\lesssim 1cm$) whiskers can be significantly moved by  rather modest air currents of the order of 1 mm/s. On the other hand, thicker ($d\sim 10$ $\mu$m) whiskers require unrealistically strong winds $v_a\gtrsim 10^4$ cm/s to be moved.
%Using the above listed parameters and setting $d\sim 10$ $\mu$m and $l/d\sim 10^2-10^3$ yields $v_a\sim 5\cdot (10^3-10^6)$ cm/s.

Table \ref{tab:air} below shows for comparison some realistic figures for the air flow velocities in the typical lab settings. They all are sufficient to create  movements of thin (or constricted) metal whiskers. This is further illustrated in Table \ref{tab:num} below.
\begin{table}[h,t]
\caption{The characteristic air flow velocities due to various sources.}
\begin{tabular}{|l|c|}
   \hline
  Air flow cause & Air flow velocity (cm/s)\footnotemark[1] \\ \hline \hline
    Room A/C & 400 \\
   HVAC Vent & 220 \\
  Person walking & 180 \\
  Door opening & 120 \\
  Expiratory air flow \footnotemark[2] & 100 \\
  Diffuser vent & 25 \\
   \hline
\end{tabular}
\footnotetext[1]{Data from Ref. \onlinecite{muller2001}, except the expiratory mode, for which the velocity was estimated assuming the average lungs expiratory volume \cite{lungs} of $\sim 1$ L and the lips opening area $\sim 1$ cm$^2$ held for 10 s.}\footnotetext[2]{Also mentioned in the NASA whisker database \cite{nasa1} as the ``gentle air flow" produced by pursing one's lips and gently blowing in the direction of the whisker. \cite{leidecker2015}}
\label{tab:air}\end{table}

\section{Brownian movements}\label{sec:brown}

Here, we consider the Brownian movements of whiskers starting with the classical analysis for a suspended particle in a fluid or gas of viscosity $\eta$. The rationale for this consideration is that the reported chaotic whisker movements seem visually similar to the Brownian patterns, and that whisker diameters are in the range of the typical Brownian particle dimensions.

\subsection{Classical Brownian movements}\label{sec:QACB}

The analysis is based on the characteristic relaxation time $\tau _\eta$ of a particle of diameter $d$ and material density $\rho\sim m/d^3$. The viscous drag force $F\sim \eta vd$ alters the particle momentum by $mv$ over time $\tau _\eta$. From that, one estimates,
\begin{equation}\label{eq:tau}
\tau _\eta\sim\frac{m}{\eta d}\sim \frac{\rho d^2}{\eta}.
\end{equation}
$\tau _\eta$ escribes the time interval during which a particle maintains a coherent translational motion.

A particle trajectory is then considered as consisting of mutually independent steps of duration $\tau _\eta$ each.
A one step displacement,
\begin{equation}\label{eq:onestep}
a\sim v_T\tau _\eta\sim \sqrt{\frac{k_BT}{m}}\tau _\eta\sim\frac{\sqrt{k_BT\rho d}}{\eta}
\end{equation}
where we have used the characteristic thermal velocity $v_T\sim\sqrt{k_BT/m}$,  $k_B$ is the Boltzmann's constant, and $T$ is the temperature.
During time $t\gg \tau _\eta$, the particle makes $t/\tau _\eta\gg 1$ random steps resulting in the average displacement
\begin{equation}\label{eq:classB}
L\sim a\sqrt{\frac{t}{\tau _\eta}}\sim \sqrt{\frac{k_BT}{\eta d}t}\equiv\sqrt{D_dt}\quad {\rm with}\quad D_d\sim\frac{k_BT}{\eta d}.
\end{equation}
Here $D_d$ is the particle diffusion coefficient.

To the accuracy of numerical multipliers, Eq. (\ref{eq:classB}) coincides with the known results for Brownian movements.\cite{snook2007,kampen2005,chandrasekhar1943,coffrey2004} As yet another illustration of this paper `order-of-magnitude' approach, note that the exact result for the Brownian movement amplitude is different by the numerical coefficient $\approx 0.5$ from that in Eq. (\ref{eq:classB}).

The characteristic amplitudes of Brownian motion are rather small, below the naked eye range. Indeed, using the viscosities  of air and water to be respectively $\eta\sim 10^{-4}$ g/cm-s and  $\eta\sim 10^{-2}$ g/cm-s, yields respectively $\tau _\eta\sim 10^{-3}$ s and  $\tau _\eta\sim 10^{-5}$ s, and the diffusion coefficients $D_d\sim 10^{-6}$ cm$^2$/s and $D_d\sim 10^{-4}$ cm$^2$/s. On the typical observation time scale $t\sim 1$ s, for a micron sized particles, $d\sim 1$ $\mu$m, the characteristic displacements will be respectively $L\sim 1$ $\mu$m and $L\sim 10$ $\mu$m, in agreement with the results of optical microscopy. \cite{brown}

The concept of classical  Brownian movements is significantly modified by the presence of external fields or constraints. \cite{snook2007,kampen2005,chandrasekhar1943,coffrey2004} In particular, for the case of a harmonic oscillator, the classical result $L\propto\sqrt{t}$ holds for relatively short displacements $L<L_T\sim\sqrt{k_BT/K}$, where $L_T$ is the time independent equipartition value limited by the restoring force $F=-KL$.

\subsection{Brownian movements of metal whiskers}\label{sec:QABW}

Along the same lines, one can consider random whisker bending over angle $\phi$ during time $t$ as a succession of $t/\tau _\eta\gg 1$ events, each producing the characteristic bending angle $\Delta \phi\sim \omega _T\tau _\eta$ where $\omega _T$ is the thermal angular velocity and $\tau _\eta$ is the coherency time.

Unlike the classical Brownian motion, the stochastic displacements here are limited  by the elastic bond setting the maximum `equipartition' value of whisker deflection $\lambda _T$. This limitation is similar to that of Brownian movements of a harmonic oscillator mentioned at the end of Sec. \ref{sec:QACB}.  Using the equipartition theorem in the form of $K_b\lambda _T ^2\sim k_BT$, yields the deflection $\lambda _T$ and its corresponding angle $\Delta\phi =\lambda _T/l$,
\begin{equation}\label{eq:Dphi}
\Delta\phi\sim\sqrt{\frac{k_BTl}{Y d^4}}.\end{equation}
The characteristic thermal angular velocity corresponding to the equipartition theorem $\omega _T\sim \sqrt{k_BT/I_0} $ is self-consistently related to $\Delta\phi$,
\begin{equation}\label{eq:selfcon}\Delta\phi \sim \omega _T/\omega _b.\end{equation}

Note that Eq. (\ref{eq:Dphi}) does not imply that the fluctuation $\Delta\phi$ is produced by a `rigid body' movement of a whisker as a whole. The same result can be obtained by considering a whisker made of mutually independent small domains of length $\delta l\ll l$ each. A single domain bending  dispersion is similar to that in Eq. (\ref{eq:Dphi}), $(\delta \phi )^2=(kT/Yd^4)\delta l$.  Multiplying $(\delta \phi )^2$ by the number of such domains $l/\delta l$ gives the dispersion $\Delta \phi$ that reproduces Eq. (\ref{eq:Dphi}).

In connection with the latter paragraph, note that the characteristic  angular velocities of individual domains are by the factor $(l/\delta l)^{3/2}\gg 1$ greater than the integral angular velocity $\omega _T$. Hence, the latter derivation implies high frequency chaotic dynamics with rapidly varying local configurations limited to the integral angle of Eq. (\ref{eq:Dphi}). In fact, Eq. (\ref{eq:Dphi}) reproduces the classical result for fluctuations in curvature of long molecules. \cite{bresler1939}

We now estimate the characteristic time of establishing the above maximum deflection. Similar to the treatment of classical Brownian movements in Sec. \ref{sec:QACB}, the angle of deflection over time $t$ is accumulated via $t/\tau _\eta$ mutually independent characteristic partial deflections $\delta \phi\sim\omega _T\tau _\eta$, i. e.
\begin{equation}\label{eq:Dphi1}\Delta\phi\sim \sqrt{\frac{t}{\tau _\eta}}\omega _T\tau _\eta.\end{equation}

To evaluate $\tau _\eta$, we use the Newton's second law presenting the viscous drag of Eq. (\ref{eq:viscdrag}) in the form $F\sim mv/\tau _\eta$. This yields $\tau _\eta\sim  \rho d^2/\eta$, which coincides with the estimate for a sphere of diameter $d$ in Eq. (\ref{eq:tau}) (again, to numerical factors).

Combining Eqs. (\ref{eq:Dphi}) and (\ref{eq:Dphi1}) and substituting $\tau _\eta$ yields the characteristic time of reaching the maximum bending $\Delta\phi$ [of Eq. (\ref{eq:Dphi})],
\begin{equation} \label{eq:tD} t_{\Delta\phi}\sim \frac{\eta}{Y}\left(\frac{l}{d}\right)^4. \end{equation}

Numerical estimates in Table \ref{tab:num} below show that, for realistic parameters, the predicted Brownian bending angle $\Delta \phi$ is too small to explain the observed whisker movements, although possible bending times $t_{\Delta\phi}$ of the order of  seconds are compatible with the data. Even for severely constricted whiskers with local diameter $d\sim 0.1$ $\mu$m, the predicted bending angle $\Delta \phi\propto d^{-2}$ remains small compared to the observations.

\subsection{Brownian movements vs. viscous drag}\label{sec:brovis}
A comment is in order regarding one intuitively justified hypothesis, that spontaneous whisker movements could be of the same nature as that of the often observed stumbling of small dust/smoke particles in the air,\cite{wikibrown} both being Brownian. Here we briefly discuss how Brownian movements are typically much weaker than that due to viscous drag, and are hardly relevant for metal whiskers.

The ratio of the characteristic displacements due to viscous drag, $v_at$ over that due to Brownian diffusion, $\sqrt{D_dt}$ can be expressed in terms of microscopic parameters if we take into account the definition for $D_d$ in Eq. (\ref{eq:classB}) along with the standard representation for viscosity, $\eta\sim nk_BTl_0/v_T$ where $v_T=\sqrt{k_BT/m_0}$ is the thermal velocity, $m_0$ is the molecular mass, $n$ is the molecule concentration, and $l_0=1/(n\sigma)$ is the molecule mean free path with $\sigma$ being the molecular scattering cross section. This yields,
\begin{equation}\label{eq:relcont}
\frac{v_at}{\sqrt{D_dt}}\sim \frac{v_a}{v_T}\sqrt{\frac{v_Ttd}{\sigma}}.\end{equation}

The right-hand-side in Eq. (\ref{eq:relcont}) is readily estimated. We take a modest $v_a\sim 1$ mm/s, the typical $v_T\sim 10^5$ cm/s and $\sigma\sim 10^{-16}$ cm$^2$, also setting  $d\sim 1$ $\mu$m, and $t\sim 1$ s corresponding to the naked eye observation times. As a result, one gets, $v_at/\sqrt{D_dt}\sim 300$. The viscous drag will be yet more significant for bigger particles and stronger air currents.

It may be worth pointing at a more intuitive interpretation of the above. Presenting $1/\sigma =l_0n$ translates the square root in Eq. (\ref{eq:relcont}) into the form $\sqrt{N}\equiv \sqrt{nl_0vtd}$ with $N\gg 1$. Therefore, this is a very large number $N\gg 1$ of molecules involved that makes the viscous drag more efficient by the factor proportional to $\sqrt{N}$.

\begin{figure}[th]
\includegraphics[width=0.37\textwidth]{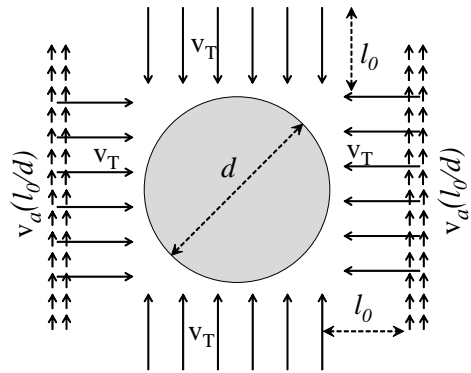}
\caption{A discrete (4-side) sketch of a particle bombarded by molecules having on average the thermal velocity $v_T$  and capable of colliding with the particle when they are within the range of the mean free path $l_0$ from it. Shown in short arrows are vertical (in the plane of figure) air currents of low velocity $v_a$ that is reduced by the factor $l_0/d$ in the domain of molecules  capable of colliding with the particle.  \label{Fig:brovis}}
\end{figure}

The two mechanisms are additionally illustrated in Fig. \ref{Fig:brovis} where a particle is bombarded with molecules within a layer of thickness $l_0$. We consider the particle movements in the vertical direction, parallel to which, there is a flow of air with velocity $v_a$ at distance $\sim d$ in horizonal direction ($d$ is the characteristic distance over which the air flow is disturbed by the particle). The airflow velocity at distance $l_0$ from the particle is estimated as $\sim v_al_0/d$. Each molecule of mass $m_0$ hitting the particle in horizontal direction, transfers the momentum $m_0v_al_0/d$, and the momenta transferred from the left and right do not cancel each other. The momentum transferred over time $t$ is estimated as $$\Delta p_{\rm visc}\sim nv_TAtm_0v_al_0/d$$ where $A$ is the particle area. Dividing the latter by time $t$ results in the viscous drag force $\eta v_ad$ with the standard viscosity $\eta =nl_0k_BT/v_T$. To the contrary, for the Brownian movements, the momenta transferred from the top and bottom collisions mostly cancel each other and only the $1/\sqrt{N}$ fraction of them remains unbalanced as a fluctuation (we skip the details of derivation).

Our conclusion here is that the Brownian motion is hardly observable with metal whiskers. Similarly, the most visible movements of dust particles in the air, are often explained by tiny air currents due to thermal convection or other ill controlled factors.

\section{Mechanical models}\label{sec:mech}
\subsection{Sample vibrations}\label{sec:vib}
Here we test the hypothesis that metal whisker movements are caused by uncontrolled vibrations of the film transmitted from  remote sources, such as some power equipment, transportation, etc.

\begin{figure}[th]
\includegraphics[width=0.47\textwidth]{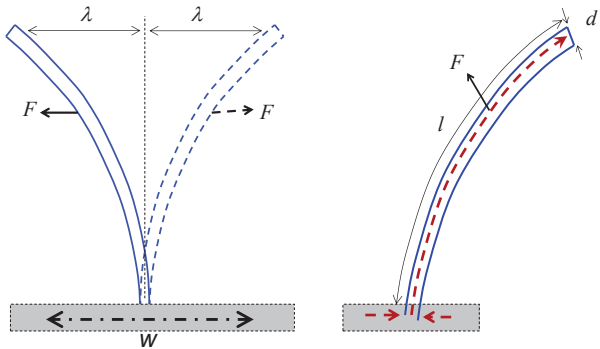}
\caption{Left: whisker movements due to the inertial (fictitious) forces caused by the sample mechanical vibrations with acceleration $w$. Dash-dot arrow illustrated the sample acceleration. Right: whisker deflection caused by the momentum change due to its growth (garden hose model). The dashed arrows represent the material movement to the base of the whisker and its caused movement (growth) of a whisker as a whole. \label{Fig:mech}}
\end{figure}

The characteristic deflection caused by such vibrations can be estimated by using the non-inertial frame of reference pinned to the sample, as illustrated in Fig. \ref{Fig:mech}. In that frame, there will be the inertial (fictitious) force $F_i=-mw$ exerted on the whisker, where $w$ is the sample acceleration relative to the laboratory system. Substituting that force in Eq. (\ref{eq:deflection}) and using $m\sim \rho d^2l$, yields
\begin{equation}\label{eq:inert}
\lambda \sim\frac{\rho d^2w}{Y}\left(\frac{l}{d}\right)^4.
\end{equation}

For numerical estimates in Table \ref{tab:num}, we used the extremum value related to acceleration due to gravity, $w=g=9.8$ m/s$^2$ (that can be felt in a roughly moving truck \cite{brandt2011}). Both that and a `slightly rough' \cite{gidhar2004} acceleration $w\sim 0.1g$, are well above the standard lab exposed `smooth' accelerations, which fall in the range \cite{gidhar2004}  of $w<0.01g$. For the latter,  the whisker deflection are at least by the factor of 0.01 smaller than shown in Table \ref{tab:num} (for $w=g$), and they are much smaller than observed.  However, severely constricted whiskers with $d\sim 0.1$ $\mu$m can respond by deflections that are by two orders of magnitude higher than given in Table \ref{tab:num}. They would be consistent with the observed movements.

A word of caution is in order regarding possible resonance phenomena amplifying the vibrational amplitudes for frequencies in the observed whisker motion range, say $\omega _b\sim 30$ in Table \ref{tab:num}. While the possibility of such amplification depends on the spectral composition of a noise, the resonance amplification factor for whiskers can be significant, $\omega\tau _r\gg 1$ where the velocity relaxation time due to viscous friction is estimated as $\tau _r\sim m/(\eta d)$. Using the numerical parameters from Table \ref{tab:param} yields $\omega\tau _r\sim 10-100$.

\subsection{Garden hose instability}\label{sec:garden}
The fluid-conveying pipe, particularly a garden hose can exhibit oscillating behavior due to interactions between the moving fluid and a hose. Various mechanisms of such interactions have been considered (see Refs. \onlinecite{doare2002, kuiper2008} and references therein). Here we concentrate on one of them applicable to the very slow material flow relevant for the case of growing whiskers.

Denoting $v_w=dl/dt$ the time independent characteristic whisker growth, the corresponding momentum transfer is due to the change in whisker mass,
\begin{equation}\label{eq:momtr}
\frac{\partial p}{\partial t}=v_w\frac{\partial m}{\partial t}\sim v_w\rho d^2\frac{\partial l}{\partial t}=\rho v_w^2d^2.\end{equation}

Assuming the whisker bending angle ($\sim 90^o$) such that the entire force $F=\partial p/\partial t$ is directed more or less perpendicular to the whisker, and substituting that force in Eq. (\ref{eq:deflection}) yields the relative whisker deflection,
\begin{equation}\label{eq:reldef}
\frac{\lambda}{l}\sim\left(\frac{v_w}{v_s}\right)^2\left(\frac{l}{d}\right)^2.\end{equation}

The latter ratio is extremely small because of the relatively low whisker growth velocity $v_w\sim 1$ {\AA}/s compared to the sound velocity $v_s\sim 10^5$ cm/s. As seen from Table \ref{tab:num}, this mechanism should be totally rejected even for severely constricted whiskers.

\begin{figure}[t!]
\includegraphics[width=0.47\textwidth]{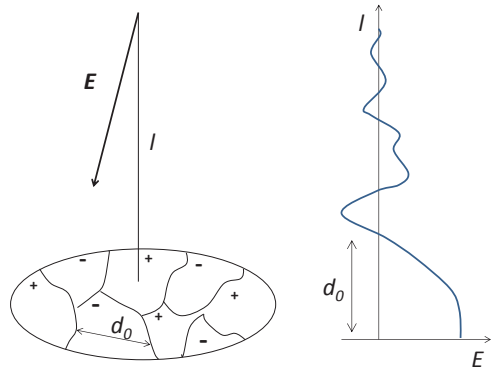}
\caption{Charge patch model. Left: positive and negative charge patches of characteristic linear dimension $d_0$ generating random electric field vectors at distances $l\gg d_0$. Right: electric field fluctuations with absolute value decaying as $E\propto 1/l$ with distance $l\gg d_0$ from the surface. \label{Fig:patch}}
\end{figure}

\section{Electric forces}\label{sec:electric}

Here we consider a hypothesis that whisker movements are caused by interactions of whiskers with charged patches on the underlying surface.

\subsection{Charge patch model and static deflection}\label{sec:CPM}

The charge patch model \cite{karpov2014,karpov2015} takes into account local electric charges in the metal surface originating from various imperfections: grain boundaries, local deformations, variations in chemical composition, contaminations, dislocations, etc. It was able to describe many properties of whiskers, such as the existence of dormant and linear growth stages, their high aspect ratios, growth velocities, versatility of factors affecting whisker appearances and kinetics, and broad statistical distributions of whisker parameters. That model however calls upon further verifications. Earlier evidence of electric bias effects on whiskers did not discriminate between the field and current effects. \cite{liu2004,crandall} However, movements of metal whiskers caused by the electric field was experimentally demonstrated, \cite{kadesch2001} and there is a preliminary evidence that the electric field strongly accelerates whisker growth. \cite{vasko2015,umalas2015}

The charge patch model assumes relatively small surface regions illustrated in Fig. \ref{Fig:patch} with characteristic linear dimension $d_0\lesssim 10$ $\mu$m where individual patch charges are mutually uncorrelated. In the near surface region $l< d_0$, the surface field is more or less uniform and perpendicular to the surface; its absolute value $E_0$ is determined by the local charge density. Farther from the surface, at $l\gg d_0$, the oppositely charged random patches mostly cancel each other contributions making the field decay as $E\sim E_0d_0/l$ on average. In that latter region, the field vectors are oriented almost randomly. The characteristic maximum surface field at $l< d_0$ was estimated to fall in the range of $E_0\sim 10^4-10^6$ V/cm. \cite{karpov2014,karpov2015}

For a whisker of length $l$ at angle $\theta$ with the normal, the characteristic force parallel to the surface can be estimated as $F\sim (pE/l)\cos\theta\sim (pE_0d_0/l^2)\cos\theta$ where the whisker dipole moment is $p=\beta E$ and its polarizability  $\beta$ is estimated as that of a long thin metal rod, \cite{landau1984} i. e. $\beta\sim l^3$. Substituting that force in Eq. (\ref{eq:deflection}) yields
\begin{equation}\label{eq:elect}
\frac{\lambda}{l}\sim\frac{E_0^2}{Y}\left(\frac{d_0}{l}\right)^2\left(\frac{l}{d}\right)^4\cos\theta.
\end{equation}

Numerically, the latter ratio can be comparable to (or even formally exceed) unity for reasonable parameter values as illustrated in Sec. \ref{sec:num}. Therefore, the electrostatic forces due to charge patches are strong enough to cause a significant static whisker bending with $\lambda /l\lesssim 1$.

\subsection{Ion dipoles}\label{sec:idip}
To estimate the possibility of whisker movements, a mechanism of temporal variations in patch charges is needed. Here we follow the concept of electric field fluctuations due to the time dependent electric dipoles on metal surfaces (or in its underlying layers). \cite{safavi2011,daniilidis2011}

The dipoles are caused by ions diffusing across the surface. Each diffusing ion of charge $e$ creates a mean square dipole fluctuation $(\delta p)^2\sim e^2D_it$ where $D_i$ is the ion diffusion coefficient. The corresponding dispersion in electric fields at distance $l$ from the surface becomes
\begin{equation}\label{eq:fluctfield}
(\delta E)^2\sim \frac{(nl^2)e^2D_it}{l^6}.\end{equation}
In Eq. (\ref{eq:fluctfield}), we have taken into account the dipoles in the area $\sim nl^2$ where $n$ is the dipole concentration equal to that of the surface ions. This leads to the estimate,
\begin{equation}\label{eq:deltaE}
\frac{\delta E}{E}\sim \sqrt{\frac{D_it}{nl^2d_0^2}}.\end{equation}

The relative whisker movement amplitudes $\lambda /l$ are proportional to the above $\delta E/E$. As is numerically demonstrated in Sec. \ref{sec:num}, that ratio is too small to explain the observable whisker movements. That smallness is due to extremely slow ion diffusion. Even taking ten orders of magnitude higher diffusion coefficient (along grain boundaries \cite{su2011}) than in Table \ref{tab:param} retains that smallness.

\subsection{Non-equilibrium charges}\label{sec:noneq}

\begin{figure*}
\centering
\includegraphics[width=0.45\textwidth]{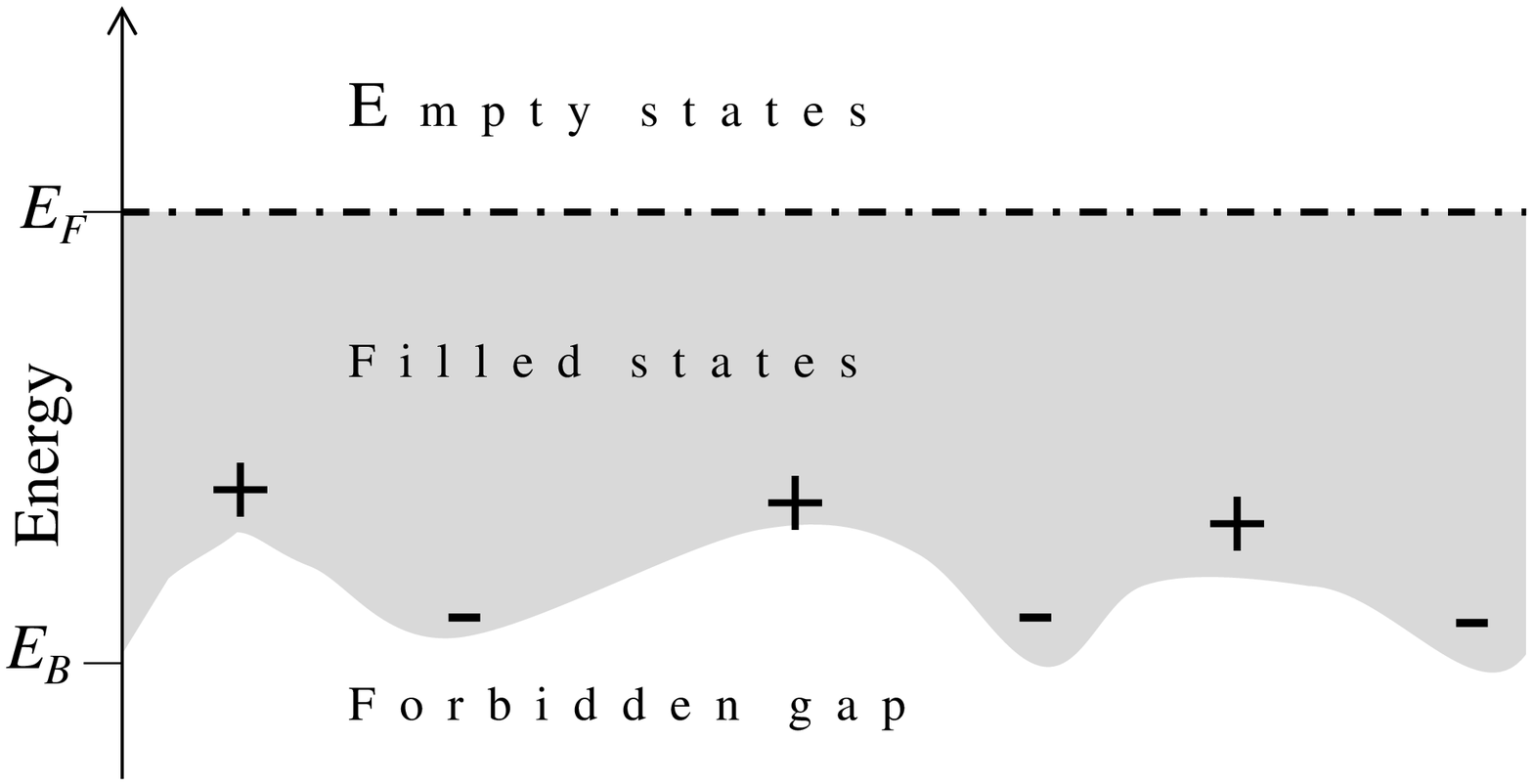}\includegraphics[width=0.45\textwidth]{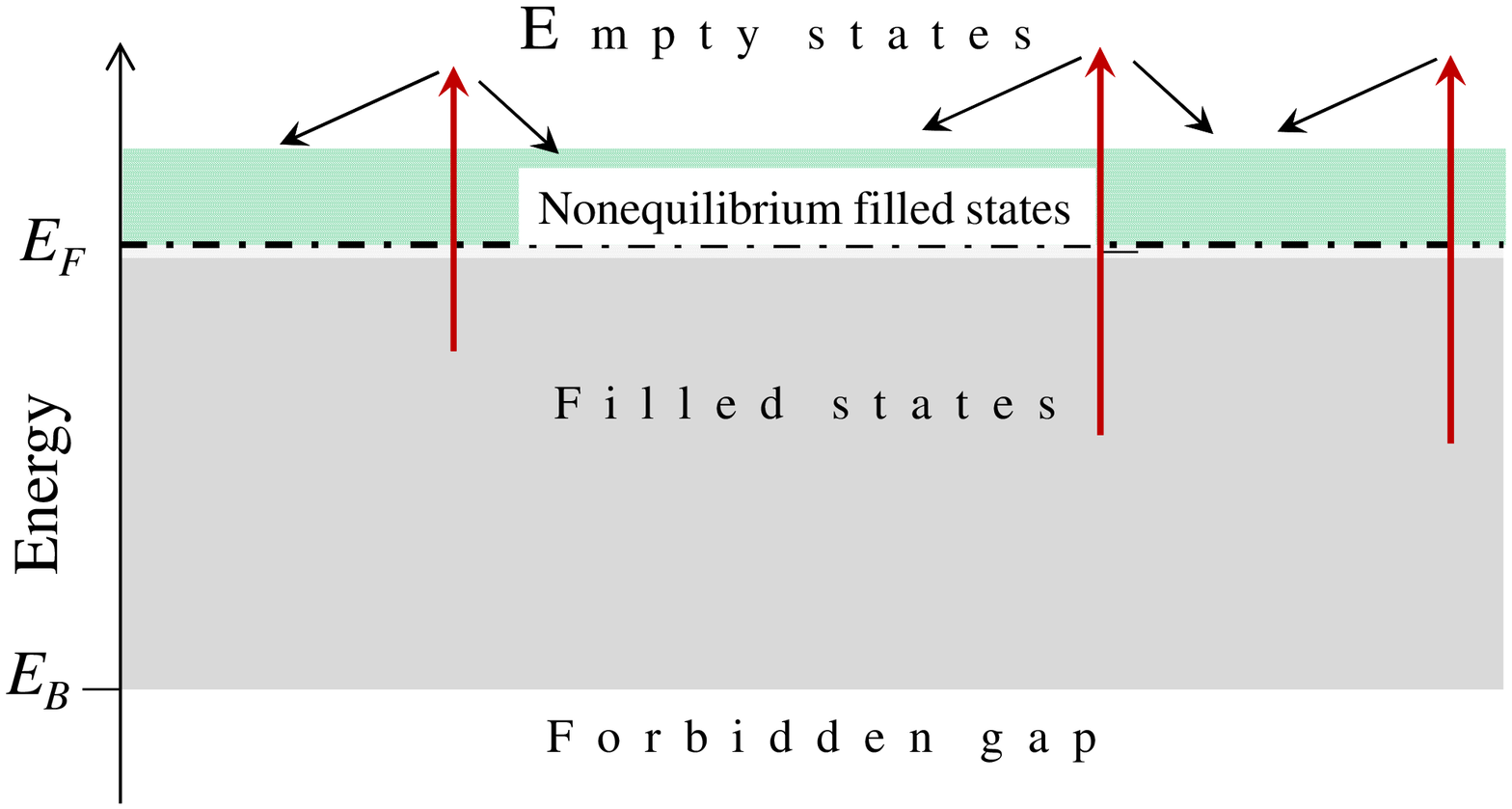}
\caption{\label{Fig:Fermi} Energy band diagram for electrons in a nonuniform metal. $E_F$ is the Fermi level separating in equilibrium the filled (gray area) and empty states. Left: Equilibrium state. The bottom edge of the partially filled band is curved due to the random fields in the system. The regions of negative and positive curvature correspond respectively to positively and negatively charged local regions. Right: Nonequilibrium state under light (or radiation) exposure. The vertical and tilted arrows show respectively the processes of photo-excitation and electron drift. The latter processes generate the nonequilibrium population of the electrons above the Fermi level thus leveling out the local electric field distribution.}
\end{figure*}

Some external perturbations, such as light or ionizing radiation, create nonequilibrium electron distributions. The nonequilibrium electrons drift in the local electric fields screening them as illustrated in Fig. \ref{Fig:Fermi}. These temporal variations in local electric fields can affect whisker configurations, which constitutes yet another candidate mechanism of whisker movements. It is important that such electric currents are much faster than that due to the above considered ion diffusion because the electron diffusion coefficients are by up to $\sim 20$ orders of magnitude higher.

Note that the contact with a substrate will additionally enhance the above mechanism because the latter will partially absorb either electrons or holes (depending on the chemical potentials of a metal film and its substrate) thus increasing the electric nonuniformities in the film. Therefore, that mechanism is predicted to be substrate dependent.

The problem of photoinduced response of imperfect metal surfaces apparently has not been systematically addressed. Given a large number of imperfections in the metal grain boundaries, oxides, etc., the time scale of establishing new steady state distribution of electrons can vary by many orders of magnitude between different metal recipes and even between different specimens of the same recipe. Lacking metal related data one can look into the photoinduced recharging data for semiconductors, especially semiconductor polycrystalline  thin films, which are ample in the literature for the past decades. In particular, the phenomena of surface photoconductivity and persistent photoconductivity have been studies for multiple semiconductor films (see multiple references in Refs. \onlinecite{lang1977,tarun2013,bhatnagar2014}.  In may cases, the photoinduced changes in charge density were found to be much greater than the corresponding equilibrium values; both positive and negative signals have been observed in different materials.

The characteristic equilibration times for photoinduced charge transfer in semiconductors vary between different systems from microseconds to days. Assuming similar amplitudes and equilibration times for charges photoinduced in imperfect metal surfaces, Eq. (\ref{eq:elect}) predicts a possibility of `spontaneous' metal whisker movements strongly varying between different specimens and light intensity, and ambient conditions.

A serious problem with the latter explanation is that it is difficult to make it more quantitative for the same reasons as it is for the case of semiconductors where all kinds of imperfections are involved in recharging phenomena. However, that explanation has the advantage of offering direct experimental verifications by changing the light intensity and temperature while trying to observe whisker movements. Another attractive feature is that it helps to reconcile various seemingly conflicting testimonies, some of which indicate `spontaneous' whisker movements, while others do not.

We shall end this section with one rather arbitrary estimate that shows how given plausible parameters of the metal surface, the above discussed mechanism can predict significant whisker movements. We proceed from Eq. (\ref{eq:deltaE}) with the following modifications: (i) replace the ionic diffusion coefficient with the electronic one, $D_i\rightarrow D_e$, and (ii) multiply the result on the right-hand-side of Eq. (\ref{eq:deltaE}) with the ratio $\delta n/n$ of the photoinduced electron concentration change over its average equilibrium value. This yields
\begin{equation}\label{eq:deltaE1}
\frac{\delta E}{E}\sim \frac{\delta n}{n}\sqrt{\frac{k_BT\mu _et}{nel^2d_0^2}}.\end{equation}
Here we have taken into account the Einstein relationship between the mobility and diffusivity, $\mu _e=(e/k_BT)D_e$. As seen from the numerical estimates in Table \ref{tab:num}, this mechanism appears plausible (note that $\delta n/n$ can be larger than one in absolute value based on the data for photoconductivity in semiconductors).

\section{Numerical Estimates}\label{sec:num}

\begin{table}[th]
\caption{Whisker related geometrical and material parameters.}
\begin{tabular}{|l|c|c|c|}
   \hline
  Parameter & Value  & Ref.  \\   \hline
    density $\rho$ g/cm$^3$  & 7.4 (7.1)\footnotemark[1] &  \onlinecite{matpar}  \\ \hline
  Young's modulus $Y$ GPa& 50 (108) & \onlinecite{matpar} \\ \hline
  sound velocity, $v_s$,$10^{5}$ cm/s & 2.7  (3.9) & \onlinecite{matpar} \\ \hline
  diameter,\footnotemark[2] $d$ $\mu$m & 0.1-20 &\onlinecite{brusse2002,davy2014,fang2006,panashchenko2009,susan2013} \\ \hline
 length, $l$  & 10$\mu$m-25 mm & \onlinecite{brusse2002,davy2014,fang2006,panashchenko2009,susan2013}  \\ \hline
   dynamic air viscosity, $\eta$ g/cm-s  & $2\cdot 10^{-4}$ &\onlinecite{air}   \\ \hline
   kinematic air  viscosity, $\eta /\rho _a$, cm$^2$/s  & 0.15 & \onlinecite{air} \\ \hline
   patch size, $d_0$ $\mu$m & 0.1-10 & \onlinecite{karpov2014,karpov2015}\\ \hline
   electric charge density, $n$ $e$/cm$^2$&$10^{10}-10^{12}$  & \onlinecite{karpov2014,karpov2015} \\ \hline
   near surface field, $E_0$ V/cm, &$10^{4}-10^{6}$  & \onlinecite{karpov2014,karpov2015} \\ \hline
   ion diffusion coefficient $D_i$, cm$^2$/s, &$10^{-16}$  & \onlinecite{woodrow2006} \\ \hline
   electron mobility $\mu _e$, cm$^2$/V-s, &1000 & \onlinecite{sze}\\\hline

\end{tabular}
\footnotetext[1]{The numbers in the top three rows are for Sn; numbers in parentheses are for Zn.}
\footnotetext[2]{Whisker diameters and lengths are mutually uncorrelated and are described by log-normal probabilistic distributions. \cite{davy2014,fang2006,panashchenko2009} In this paper, we use four pairs of values for $d$ and $l$ (Table \ref{tab:num}) to illustrate the scale of variations in their depending quantities. }
\label{tab:param}\end{table}

\begin{table}[thb]
\caption{Characteristic whisker vibration parameters for four different geometries.}
\begin{tabular}{|l|c|c|c|c|}
   \hline
  & $d=1$$\mu$m  & $d=1$$\mu$m & $d=10$$\mu$m & $d=10$$\mu$m\\
   Quantity & $l=1$mm & $l=10$mm & $l=1$mm & $l=10$mm \\ \hline \hline
    $v_a$cm/s,  & $5\cdot 10^2$ &  $0.05$  & $5\cdot 10^6$  & $5\cdot 10^3$ \\
    Eq. (\ref{eq:wind}) &  &  &  & \\
    $Re $\footnotemark[1] &   500 & 0.5 & $5\cdot 10^6$ & $5\cdot 10^4$\\ \hline
  $\omega _b$, s$^{-1}$& 3000 & 30 &30,000  &300 \\
  Eq. (\ref{eq:freq}) &  &  &  & \\\hline
  $\Delta\phi$, rad & $10^{-4}$ & $3\cdot 10^{-4}$ & $
  10^{-6}$ & $3\cdot 10^{-6}$ \\
  Eq.  (\ref{eq:Dphi1}) &  &  &  & \\\hline
 $t_{\Delta\phi}$, s & $2\cdot 10^{-4}$ & 2 & $2\cdot 10^{-8}$ & $2\cdot 10^{-4}$\\
  Eq.  (\ref{eq:tD}) &  &  &  &   \\ \hline
 $\lambda/l$,Eq. (\ref{eq:inert}) & $2\cdot 10^{-3}$ & 1 & $2\cdot 10^{-5}$ & 0.02   \\
 for $w=g$ &  &  &  & \\
   \hline
   $\lambda/l$,Eq. (\ref{eq:reldef})  & $10^{-7}$ &$10^{-5}$  & $10^{-9}$ & $10^{-7}$ \\ \hline
   $\lambda/l$,Eq. (\ref{eq:elect}),\footnotemark[2]  & $10^{-3}-10$ &$0.1-10^3$  & $10^{-7}-10^{-3}$ & $10-10^{5}$ \\
   $\cos \theta \sim 1$ &  &  &  &\\ \hline
   $\delta E/E$,  & $10^{-9}$ &$10^{-10}$  & $10^{-9}$ & $10^{-10}$ \\
   Eq. (\ref{eq:deltaE})\footnotemark[3] &  &  &  & \\\hline
   $\delta E/E$,  & $0.4$ &$0.04$  & $0.4$ & $0.04$ \\
   Eq. (\ref{eq:deltaE1})\footnotemark[3] &  &  &  & \\\
   with $\delta n/n=1$ &  &  &  &\\ \hline
\end{tabular}
\footnotetext[1]{$Re=lv_a\rho _a/\eta$ denotes the Reynolds number.}
\footnotetext[2]{The computed ratios $\lambda /l>1$ are formal and must be replaced with 1.}
\footnotetext[3]{For the observation time $t=1$ s.}
\label{tab:num}\end{table}
Our numerical estimates are based on a set of parameters summarized in Table \ref{tab:param}. Note that the difference between the most whisker prone metals, Sn and Zn is not very significant and Sn parameters are used everywhere in Table \ref{tab:num}.

Table \ref{tab:num} contains the derivative parameters quantitatively predicting the ability of whiskers to move due to the mechanisms considered in this paper. It presents numerical estimates for several quantities defined in equations that are specified in the table. Such are the characteristic velocity of air flow capable of whisker movements, the Brownian movement whisker fluctuation angle and the characteristic time of its accumulation, relative whisker deflections due to external vibrations and garden hose effect, and relative field fluctuation due to the time dependent ion dipoles, and non-equilibrium electrons.

The relations between these predictions and the whisker movement observations were stated in for each of the mechanisms in the body of text accompanying the equations specified in Table \ref{tab:num}.

\section{Conclusions}\label{sec:concl}

The numerical estimates of whisker vibrational parameters in Table \ref{tab:num} point at the following. \\

1) Minute air currents can easily be responsible for the movements of thin ($d\lesssim 1$ $\mu$m) whiskers, while movements of thick whiskers ($d\gtrsim 10$ $\mu$m) require much stronger, even turbulent ($Re >10^4$) air flows.\\
2) It is much less likely (if possible at all for maybe especially constricted whiskers) that the Brownian motion can be responsible for the observed whisker vibrations.\\
3) External mechanical vibrations from power equipment etc. can be responsible for whisker movements in a rather harsh environment ranging between tracks moving on rural roads to labs under construction; again, the case of severely constricted whiskers make it more likely.\\
4) The garden hose instability can never be a source of whisker movements.\\
5) Electric forces can be responsible for whisker movements if their temporal dependence is determined by the electronic (as opposed to the very slow ionic) diffusion. That latter possibility can explain the observations of `spontaneous' whisker movements, and it offers direct experimental verifications via various dark-light conditions.

Certain experiments could help to verify the above conclusions. In particular, observing whisker dynamics vs. (a) light intensity, (b) ionizing radiation, (c) air ionizer/deionizer action, and (d) ambient viscosity ranging from that of transparent liquids to vacuum conditions. \cite{gordon2015}

As another useful outcome, the above results open a venue to more quantitative discussions of whisker properties. For example, they show how to predict whisker behavior on accelerated platforms (say, in auto-industry), or how to make simple quantitative estimates of whicker properties observing their movements under air flows.

At the end, the question of well enough documented observations of `spontaneous' whisker movements calls upon further attention, as opposed to the observations of air flow effects that have found direct quantitative confirmation in the above. The possibility of modifying the whisker growing surfaces by shining light or ionizing radiation may have practical applications.

\section*{Acknowledgement}
The author is grateful to D. Shvydka, G. Davy, B. Rollins, A. D. Kostic, J. Brusse, H. Leidecker, L. Panashchenko, S. Smith, and A. V. Subashiev for useful discussions. Courtesy of the NASA Electronic Parts and Packaging (NEPP) Program is greatly appreciated.

\end{document}